\documentclass[aps,prc,nofootinbib,
twocolumn,
superscriptaddress]{revtex4-1}
\RequirePackage{lineno}
\usepackage{bbm}
\usepackage{graphicx}
\usepackage{amsmath}
\usepackage{amsfonts,amsbsy}
\usepackage{amssymb}
\usepackage{subfigure}
\usepackage{comment}
\usepackage[usenames,dvipsnames]{color}
\newcommand{\gc}{{\mathrm{ch}-\gamma}}

\newcommand{\la}{\left<}
\newcommand{\ra}{\right>}
\newcommand{\lf}{\left(}
\newcommand{\rf}{\right)}

\newcommand{\ph}{\gamma}
\newcommand{\ch}{\mathrm{ch}}

\newcommand{\nc}{\la N_\ch \ra}
\newcommand{\np}{\la N_\ph \ra}
\newcommand{\npf}{\la N_\ph(N_\ph-1) \ra}
\newcommand{\ncf}{\la N_\ch(N_\ch-1) \ra}

\newcommand{\be}{\begin{equation}}
\newcommand{\ee}{\end{equation}}
\newcommand{\bea}{\begin{eqnarray}}
\newcommand{\eea}{\end{eqnarray}}
\newcommand{\ndyn}{\nu_{\mathrm{dyn}}}
\newcommand{\rmn}{r_{m,1}}
\newcommand{\corr}{\mathrm{corr}_{\gc}}
\newcommand{\ncp}{ \sqrt{\langle N_\ch \rangle \, \langle N_\ph \rangle}}

\newcommand{\black}{\color{black}}

\begin{document}

\title{Charged-to-neutral correlation at forward rapidity in Au+Au collisions at $\sqrt{s_{NN}}$=200 GeV}

\affiliation{AGH University of Science and Technology, Cracow, Poland}
\affiliation{Argonne National Laboratory, Argonne, Illinois 60439, USA}
\affiliation{University of Birmingham, Birmingham, United Kingdom}
\affiliation{Brookhaven National Laboratory, Upton, New York 11973, USA}
\affiliation{University of California, Berkeley, California 94720, USA}
\affiliation{University of California, Davis, California 95616, USA}
\affiliation{University of California, Los Angeles, California 90095, USA}
\affiliation{Universidade Estadual de Campinas, Sao Paulo, Brazil}
\affiliation{Central China Normal University (HZNU), Wuhan 430079, China}
\affiliation{University of Illinois at Chicago, Chicago, Illinois 60607, USA}
\affiliation{Cracow University of Technology, Cracow, Poland}
\affiliation{Creighton University, Omaha, Nebraska 68178, USA}
\affiliation{Czech Technical University in Prague, FNSPE, Prague, 115 19, Czech Republic}
\affiliation{Nuclear Physics Institute AS CR, 250 68 \v{R}e\v{z}/Prague, Czech Republic}
\affiliation{Frankfurt Institute for Advanced Studies FIAS, Germany}
\affiliation{Institute of Physics, Bhubaneswar 751005, India}
\affiliation{Indian Institute of Technology, Mumbai, India}
\affiliation{Indiana University, Bloomington, Indiana 47408, USA}
\affiliation{Alikhanov Institute for Theoretical and Experimental Physics, Moscow, Russia}
\affiliation{University of Jammu, Jammu 180001, India}
\affiliation{Joint Institute for Nuclear Research, Dubna, 141 980, Russia}
\affiliation{Kent State University, Kent, Ohio 44242, USA}
\affiliation{University of Kentucky, Lexington, Kentucky, 40506-0055, USA}
\affiliation{Korea Institute of Science and Technology Information, Daejeon, Korea}
\affiliation{Institute of Modern Physics, Lanzhou, China}
\affiliation{Lawrence Berkeley National Laboratory, Berkeley, California 94720, USA}
\affiliation{Massachusetts Institute of Technology, Cambridge, Massachusetts 02139-4307, USA}
\affiliation{Max-Planck-Institut f\"ur Physik, Munich, Germany}
\affiliation{Michigan State University, East Lansing, Michigan 48824, USA}
\affiliation{Moscow Engineering Physics Institute, Moscow Russia}
\affiliation{National Institute of Science Education and Research, Bhubaneswar 751005, India}
\affiliation{Ohio State University, Columbus, Ohio 43210, USA}
\affiliation{Old Dominion University, Norfolk, Virginia 23529, USA}
\affiliation{Institute of Nuclear Physics PAN, Cracow, Poland}
\affiliation{Panjab University, Chandigarh 160014, India}
\affiliation{Pennsylvania State University, University Park, Pennsylvania 16802, USA}
\affiliation{Institute of High Energy Physics, Protvino, Russia}
\affiliation{Purdue University, West Lafayette, Indiana 47907, USA}
\affiliation{Pusan National University, Pusan, Republic of Korea}
\affiliation{University of Rajasthan, Jaipur 302004, India}
\affiliation{Rice University, Houston, Texas 77251, USA}
\affiliation{University of Science and Technology of China, Hefei 230026, China}
\affiliation{Shandong University, Jinan, Shandong 250100, China}
\affiliation{Shanghai Institute of Applied Physics, Shanghai 201800, China}
\affiliation{SUBATECH, Nantes, France}
\affiliation{Temple University, Philadelphia, Pennsylvania 19122, USA}
\affiliation{Texas A\&M University, College Station, Texas 77843, USA}
\affiliation{University of Texas, Austin, Texas 78712, USA}
\affiliation{University of Houston, Houston, Texas 77204, USA}
\affiliation{Tsinghua University, Beijing 100084, China}
\affiliation{United States Naval Academy, Annapolis, Maryland, 21402, USA}
\affiliation{Valparaiso University, Valparaiso, Indiana 46383, USA}
\affiliation{Variable Energy Cyclotron Centre, Kolkata 700064, India}
\affiliation{Warsaw University of Technology, Warsaw, Poland}
\affiliation{University of Washington, Seattle, Washington 98195, USA}
\affiliation{Wayne State University, Detroit, Michigan 48201, USA}
\affiliation{Yale University, New Haven, Connecticut 06520, USA}
\affiliation{University of Zagreb, Zagreb, HR-10002, Croatia}

\author{L.~Adamczyk}\affiliation{AGH University of Science and Technology, Cracow, Poland}
\author{J.~K.~Adkins}\affiliation{University of Kentucky, Lexington, Kentucky, 40506-0055, USA}
\author{G.~Agakishiev}\affiliation{Joint Institute for Nuclear Research, Dubna, 141 980, Russia}
\author{M.~M.~Aggarwal}\affiliation{Panjab University, Chandigarh 160014, India}
\author{Z.~Ahammed}\affiliation{Variable Energy Cyclotron Centre, Kolkata 700064, India}
\author{I.~Alekseev}\affiliation{Alikhanov Institute for Theoretical and Experimental Physics, Moscow, Russia}
\author{J.~Alford}\affiliation{Kent State University, Kent, Ohio 44242, USA}
\author{C.~D.~Anson}\affiliation{Ohio State University, Columbus, Ohio 43210, USA}
\author{A.~Aparin}\affiliation{Joint Institute for Nuclear Research, Dubna, 141 980, Russia}
\author{D.~Arkhipkin}\affiliation{Brookhaven National Laboratory, Upton, New York 11973, USA}
% ...

\collaboration{STAR Collaboration}\noaffiliation

%
%
\begin{abstract}
 Event-by-event fluctuations of the ratio of inclusive charged to photon multiplicities at forward rapidity in Au+Au collision at $\sqrt{s_{NN}}$=200 GeV have been studied. Dominant contribution to such fluctuations is expected to come from correlated production of charged and neutral pions.
We search for evidences of dynamical fluctuations of different physical origins. Observables constructed out of moments of multiplicities are used as measures of fluctuations. Mixed events and  model calculations are used as baselines. Results are compared to the dynamical net-charge fluctuations measured in the same acceptance. A non-zero statistically significant signal of dynamical fluctuations is observed in excess to the model prediction when charged particles and photons are measured in the same acceptance. We find that, unlike dynamical net-charge fluctuation, charge-neutral fluctuation is not dominated by correlation due to particle decay.
Results are compared to the expectations based on the generic production mechanism of pions due to isospin symmetry, for which no significant ($<1\%$) deviation is observed.
\end{abstract}

\maketitle


\section{Introduction}
Heavy-ion collisions at the Relativistic Heavy Ion Collider (RHIC) provide unique opportunities for studying matter under extreme conditions. 
One of the goals is to study the properties of a strongly interacting quark gluon plasma (sQGP) via its subsequent phase transition to a hadron gas (HG)~\cite{Adams:2005dq,Back:2004je,Adcox:2004mh,Arsene:2004fa}.
The phase transition from a sQGP to a HG is associated with de-confinement transition and chiral phase transition. One of the ways in which the de-confinement transition is expected to reveal itself is via enhanced fluctuations of conserved quantities like net-charge, strangeness, and baryon number.
For observables measured in limited regions of phase space, the grand canonical
ensemble picture provides a natural description for dynamical
fluctuations of conserved quantities~\cite{Jeon:1999gr}.
The dynamical fluctuations of quantities like charged-to-neutral pion ratio is one among very few observables that are sensitive to the chiral phase transition. When the system passes from a chirally symmetric phase to a broken phase, in a scenario of rapid cooling, there could be formation of metastable domains of disoriented chiral condensate (DCC) ~\cite{Bjd, Blaizot:1992at, Rajagopal:1992qz, Rajagopal:1995bc}. Formation and decay of DCC domains could lead to a distinct distribution of the neutral pion fraction compared to that from generic production of pions under isospin symmetry \cite{Blaizot:1992at,Rajagopal:1995bc}. If this phenomenon survives the final-state interactions, it will appear as anti-correlation, between the yields of charged and neutral pions \cite{Rajagopal:1992qz}.
In heavy-ion collisions, charged and neutral particle productions are dominant in the form of charged and neutral pions. One can use inclusive charged particle multiplicity as a surrogate for charged pions and photons for the neutral pions~\cite{Adams:2005aa}. Any form of correlation between charged and neutral pions is thus expected to affect the correlation between measured charged particles ($\ch$) and photons ($\ph$).

 The generic expectation is that due to isospin symmetry, pions of different isospins would be produced in equal abundances. 
However, the formation and decay of metastable domains of DCCs would produce pions of a particular isospin, which would lead to a large deviation in the $\gc$ correlation from expectations based on the generic pion production mechanism. 
If $f$ denotes the event wise ratio of the total number of neutral pions over the total number of all pions produced in a single event, a generic production will lead to a sharply peaked distribution around $1/3$, whereas the decay of a DCC domain would exhibit a probability function described by $P(f) = 1/2\sqrt{f}$. 
This description is different from a conventional model of pion production from a locally equilibrated system undergoing hydrodynamic evolution~\cite{Rajagopal:1995bc, Rajagopal:2000yt, Randrup:1996ay,Randrup:1997kt, Randrup:1996es, Asakawa:all, Gavin:1993bs, Gavin:1993px, Gavin:1995cp, Gavin:2001uk}.
According to the theoretical predictions~\cite{Bellwied:1998yf}, the mean momentum of such pions is inversely proportional to the size of DCC domains formed. So, the detection of DCC candidates would require sensitivity to the low-momentum region of the pion spectrum.
The existence of such a phenomenon was previously investigated in heavy-ion collisions at the SPS \cite{Aggarwal:1997hd,Aggarwal:2000aw, Aggarwal:2002tf, Collaboration:2011rsa} at $\sqrt{s_{NN}}$=17.3 GeV and at Tevatron in $p+\bar{p}$ collisions by the Minimax~\cite{Brooks:1999xy, Minimax} collaboration at $\sqrt{s_{NN}}$=1.8 TeV. In both cases, possibility of large sized DCC domain formation has been excluded by the measurements.
{\color{black} Several theoretical predictions discuss that heavy-ion collisions at RHIC would be an ideal place to search for possible signals of DCC formation~\cite{Rajagopal:2000yt,Asakawa:all,Gavin:1993bs,Gavin:1993px,Gavin:1995cp,Gavin:2001uk}. However, there are varied opinions regarding  the observability of such signals~\cite{Rajagopal:1995bc, Rajagopal:2000yt, Randrup:1996ay,Randrup:1997kt, Randrup:1996es, Asakawa:all,Gavin:1993bs,Gavin:1993px,Gavin:1995cp,Gavin:2001uk}. 
Experimental measurements at RHIC on charge-neutral fluctuations, which are sensitive to such a phenomenon, can therefore shed light on the context.

 In this paper, we present the measurement of the event-by-event fluctuation and correlation of the inclusive multiplicities of charged particles and photons in the common phase space in Au+Au collisions at $\sqrt{s_{NN}}=200$ GeV obtained with the Solenoidal Tracker At RHIC (STAR) detector \cite{star_nim}.
The STAR has the capability of simultaneous measurement of charged particles and photons at both mid-rapidity and forward rapidity. Charged particles and photons can be measured using the Time Projection Chamber (TPC)~\cite{startpc_nim} and the Barrel Electromagnetic Calorimeter (BEMC)~\cite{starbemc_nim} respectively, in the pseudo-rapidity range of $|\eta|<1$. Drawback of this approach stems from the fact that the BEMC does not have the capability to detect low-momentum photons below 500 MeV.
So, for this analysis we use a combination of two forward detectors, the Photon Multiplicity Detector (PMD)~\cite{starpmd_nim} and the Forward Time Projection Chamber (FTPC)~\cite{starftpc_nim} for simultaneous measurements of photons and charged particles respectively, in the pseudo-rapidity range of $-3.7\!<\!\eta\!<\!-2.8$. This enables us to measure the event-by-event multiplicities of photons ($\ph$) with transverse momentum as low as 20 MeV/{\it c} and charged particles ($\ch$) with transverse momentum down to $150$ MeV/{\it c}.}

To measure the event-by-event $\gc$ correlation, we use observables that are constructed from moments of the charged particle and photon multiplicity distributions. In general, the observables constructed out of central moments have a dependence on effects such as detector inefficiency. However, it can been shown that for observables constructed from proper combinations of factorial moments of multiplicities, several detector effects can be minimized~\cite{Brooks:1999xy, Minimax, nudyn, dccmodel}. 
{\color{black} In this analysis, we use observables and the approach available in the literature~\cite{Brooks:1999xy, Minimax, nudyn, dccmodel} that are specifically designed to study the sensitivity of the strength of $\gc$ correlation. 
It must be noted that there are no quantitative predictions for DCC-like correlations in terms of inclusive $\gc$ correlations in the kinematic range of our measurement. So the goal of this analysis is to search for possible evidences of dynamical $\gc$ correlations, rule out the correlations coming from known sources, and to look for deviation from expectation based on a generic model of pion production.}

This paper is organized as follows. In Section \ref{detectors}, we discuss the experimental setup used in the measurement of charged particle and photon multiplicities. We briefly discuss the data set used and the reconstruction techniques in Section \ref{data_reco}, and in Section \ref{analysis}, we introduce the observables and the method of this analysis. We summarize our results in Section \ref{results} and conclude in Section \ref{summary}. 

\section{DETECTORS}
\label{detectors}
Two detectors with overlapping geometric acceptance in the forward rapidity region, the PMD and the FTPC, have been used to simultaneously measure photons and charged particles. A combination of detectors such as the Zero Degree Calorimeter (ZDC) and the Vertex Position Detector (VPD) has been used for minimum bias trigger selection and { the collision centrality is determined using the Time Projection Chamber (TPC) \cite{startpc_nim}.}
\subsection{Forward Time Projection Chambers}
The two cylindrical FTPCs extend the phase space coverage of the STAR experiment for charged particle detection. They are located on both sides of the collision point in the pseudo-rapidity range of $2.5\!<\!{\mid\!\eta\!\mid}\!<\!4.0$ and measure the charge states and the momentum of tracks. Apart from that, the FTPCs do not have any other particle identification capability. Each FTPC has a diameter and length of 75 cm and 120 cm respectively. The FTPCs have 10 rows of readout pads, which are called pad-rows. The pad-rows are further subdivided into six sectors and each sector has 160 pads. The distance of the first pad-row from the collision point is about 163 cm. Ar and CO$_2$, with a ratio of 50:50 by mass, form the active medium of the FTPC. In order to optimize available space and to cope with high particle density, the drift field in the FTPC is radial, perpendicular to the solenoidal magnetic field of the STAR magnet. With such a design, two track resolution of up to 2 mm can be achieved. It was shown in simulation \cite{star-pmd-ftpc} that approximately $6\!-\!7\%$ of all produced charged particles fall within the acceptance of one of the FTPCs. The detailed description of FTPC may be found in \cite{starftpc_nim}. In the present analysis, we refer to the FTPC in the negative pseudo-rapidity region (in the same direction as the PMD) for charged particle measurements, unless mentioned otherwise. 
\subsection{Photon Multiplicity Detector}
The PMD is a pre-shower detector designed to measure photon multiplicities in the pseudo-rapidity region of $-3.7\!<\!\eta\!<\!-2.3$. It is located 5.4 meters away from the collision point, outside the STAR magnet. The PMD consists of a highly granular (41472 cells in each plane) pre-shower plane placed behind a three radiation length lead converter. A second detector plane, the charged particle veto (CPV), identical in granularity and dimension to the pre-shower plane, is placed in front of the lead converter. Detector planes work on the principle of gas proportional counters with a sensitive medium of Ar and CO$_2$ in a 70:30 mass ratio. The photons interacting with the lead converter produce electromagnetic showers that cover several cells on the pre-shower plane, leading to a larger cluster compared to that from a charged particle. The CPV and pre-shower planes share common electronics and data acquisition system. Since photon clusters are identified from the hits in the pre-shower plane, we have used the data from the pre-shower plane in this analysis. The number of clusters and their ADC values measured by the charged particle veto plane were used to ensure the data quality.

Previous studies \cite{starpmd_nim, star-pmd-ftpc} have established that $\sim 10\%$ of all produced photons fall within the acceptance of the PMD. Photons in the kinematic region considered are predominantly (93-96$\%$) from the decay of neutral pions~\cite{star-pmd-ftpc}. PMD does not provide momentum measurement of the photons; however, any photon with transverse momentum  above 20 MeV/{\it c} is detected and counted. A detailed description of the PMD is mentioned in \cite{starpmd_nim}. 
\section{DATA RECONSTRUCTION}
\label{data_reco}
\begin{table}[htb]
\caption{Summary of different kinematic cuts used in this analysis.}
\vspace{10pt}
\centering
\begin{tabular}{ll}
\hline
\hline
Global:& $-5<V_z<5$ cm\\
\hline
FTPC:& Primary Track : number of fit points $> 5$ \\
&$-3.7\!<\!\eta\!<\!-2.8$ (Common $\eta-\phi$ with PMD)\\
&$0.15 < p_{T}< 1.5 $ GeV/{\it c}\\
&$dca< 3$ cm \\
\hline
PMD:& Cluster ADC cut $> 8 \times$ MIP ADC \\
&$-3.7\!<\!\eta\!<\!-2.8$ (Common $\eta-\phi$ with FTPC)\\
& number of cells in a cluster $>1$\\
& CPVADC/CPVcluster $\ge$ 1.8 $\times 10^2$\\
\hline
\hline
\end{tabular}
\label{tab_datacuts}
\end{table}
A total of about half a million Au+Au minimum bias events at $\sqrt{s_{NN}}=200$ GeV have been analyzed. These events were collected by the STAR experiment in 2007. The FTPCs are calibrated using a laser calibration system~\cite{starftpc_nim}. A set of criteria is imposed for the selection of a valid FTPC track. 
The criteria for a valid track are at least five FTPC hits and the distance of closest approach ($dca$) from the primary vertex, to be less than 3 cm.
The position of the primary interaction vertex is obtained via a simultaneous fit to TPC tracks (number of fit points) with at least 10 hits. Tracks with transverse momentum in the range 0.15 $<p_{\mathrm{T}}<$ 1.5 GeV/$c$ are included in this analysis. A previous study~\cite{star-pmd-ftpc} has shown that this combination of cuts significantly reduces the effect of split tracks and the background contamination coming primarily from $\gamma$ conversion. The contamination of the charged particles due to photons is below $5\%$. 
 {\color{black} Details of the procedure for PMD calibration and the extraction of
 photon clusters can be found in Ref.~\cite{starpmd_nim}.  In order to improve the purity of the photon samples, a set of strict selection criteria is used. 
A cluster is valid if the number of cells in a cluster is  $>$ 1 and the cluster signal is 8 times larger than the average response of all cells due to a minimum ionizing particle (MIP cut). This particular choice of quality cuts increases the purity of the photon samples up to $\sim70\%$ by dominantly reducing the contamination from charged particles. The remaining 30$\%$ impurity of photons can not be removed from the data sample even if a tighter cut is applied on the photon clusters.} 
 {\color{black} Pile-up events are removed by rejecting events with a ratio of combined ADC values of all clusters (CPVADC) to total number of clusters (CPVclusters) of the CPV plane to be less than 180. Details of the kinematic cuts for the selection of tracks and clusters are mentioned in Table~\ref{tab_datacuts}. }
\subsection{Centrality selection}
The centrality determination for this analysis was done 
using the minimum bias uncorrected multiplicity of charged particles in the 
pseudo-rapidity region $\mid\!\eta\!\mid$ $<$ 0.5, as measured by the 
TPC. {\color{black}  This avoids any self-correlation between the tracks used in centrality determination and those used for the $\gc$ correlation measurements, as both analyses are performed in non-overlapping rapidity ranges and with different detector components~\cite{Luo:2013bmi}.}
\section{Analysis Method}
\label{analysis}
The measurement of charged and photon multiplicities in the pseudo-rapidity interval -3.7 $< \eta <$ -2.8 are presented. 
 In order to remove event-by-event variations of the common detector acceptance, the collision vertex position  was restricted to a narrow range of  $-5<V_z<5$ cm. For similar reasons, the mixed event analysis was also performed in a fixed centrality bin and with a collision vertex bin of $\pm$5 cm.
In this section, we discuss the observables used in this analysis. Our aim is to study the $\gc$ correlation that are sensitive to different scenarios of pion production. As widely discussed in the literature, there are two possible scenarios of pion production that affect the $\gc$ correlation.
As mentioned before, the quantity of interest in such a context is the neutral pion fraction $f=N_{\pi^0}/(N_{\pi^0}+N_{\pi^\pm})$ \footnote{\color{black} $f$ is closely related to the ratio $f^\gc = 1/(1+ 2N_\ch/N_\ph)$; therefore, fluctuation of $f$ is related to fluctuation of the ratio of charged particles and photons. $\textsc{hijing}$ simulation shows $f^\gc$ is approximately $6\%$ larger than $f$ in the coverage -3.7 $< \eta <$ -2.8.} and its fluctuation. In the scenario of generic pion production, the distribution of $f$ is a sharply peaked function, which can be assumed to be a delta function at 1/3. The other scenario is the production of DCCs for which the distribution becomes $1/2\sqrt{f}$. The moments of $f$ will be very different in the two scenarios. Different moments of the fraction $f$ can be expressed in terms of observables constructed out of a proper combination of moments~\cite{dccmodel,Brooks:1999xy,Minimax} of event-by-event multiplicities of charged particles ($N_{\ch}$) and photons ($N_\ph$). The observables for charged-to-neutral fluctuations have to be insensitive to detector effects and at the same time sensitive to a rather small strength of $\gc$ correlation.
So, the idea is to use proper combination of factorial moments of multiplicities to remove the efficiency and acceptance effects \cite{dccmodel,Brooks:1999xy,Minimax} and express the observables in terms of the moments of $f$. {\color{black} Two such observables available in the literature are used in this analysis.}

\subsection{Observables}

The observable $\ndyn^\gc$, introduced in Ref.~\cite{nudyn}, is defined as
\bea
\nu_{dyn}^\gc&=&\frac{\ncf}{\nc^2}+\frac{\npf}{\np^2}-2\frac{\la N_\ch N_\ph \ra}{\nc\np}  \nonumber \\
&=& \omega_\ch + \omega_\ph -2 \, \times \mathrm{corr}_{\gc}.
\label{eq_ndyn}
\eea
{ Here $\la \cdots \ra$ corresponds to {\color{black} an average taken over all events}.} The first two terms, $\omega_\ch$ and $\omega_\ph$, are measures of individual charged particle and photon number fluctuations. The third term, $\mathrm{corr}_\gc$, corresponds to the scaled $\gc$ correlation. 
For purely statistical fluctuations, in the limit of very large multiplicity, these individual terms would become unity. However for finite multiplicity, these terms can deviate from unity even if there are no dynamical fluctuations~\cite{Voloshin:1999yf,Wang:2012bga}. So it is difficult to reach a conclusion based on the measurements of the individual terms. However, when all three terms are added up to form $\ndyn$ (Eq.\ref{eq_ndyn}), by construction, the finite multiplicity statistical fluctuations are eliminated~\cite{nudyn}.
Therefore, $\ndyn$ becomes zero for purely statistical fluctuations and non-zero only in the presence of dynamical fluctuations or correlations of different origins~\cite{nudyn,star_kpi,star_dcc,Tribedy:2012qs}. 
For further discussion, we refer to the limit $\ndyn=0$ as the Poisson limit of this observable. 

$\ndyn$ has the additional advantage of being insensitive to detector inefficiencies and acceptance effects~\cite{Voloshin:1999yf, errnudyn}. In a later section, we test this feature of $\ndyn$ by doing a mixed event analysis. Mixed events include the same acceptance and efficiency effects as data, but can only give rise to statistical fluctuations, which should be eliminated by the design of $\ndyn$.

{\black In order to interpret the results for $\ndyn$, one must understand different limits for this observable. From the construction of the observable, it is known that, for any form of dynamical correlation or fluctuation, $\ndyn$ will become non-zero. For $\ndyn^\gc$, one such source of dynamical fluctuation could be the fluctuation of the neutral pion fraction $f$. 
Use of $\ndyn$ for the study of charged-neutral correlation in the context of DCC production at RHIC was first suggested in Ref.~\cite{Gavin:2001uk}. Predictions at RHIC were made based on the measurement of neutral pions and charged pions. For the generic case, $\ndyn^{\pi^0-\pi^\pm}$ is predicted to be zero.  In case of DCC events, $\ndyn^{\pi^0-\pi^\pm}$ was predicted to become non-zero. However, the effect of neutral pion decay was not included in such calculation. It has been pointed out~\cite{dccmodel} that if the decay of neutral pions is taken into account, the observed value of $\ndyn^{\gc}$ can become non-zero, even in the generic case. This generic limit of $\ndyn^{\gc}$ is not universal and is dependent on the average multiplicity of photons and charged particles. So a deviation from the Poisson limit of $\ndyn^{\gc}$ may not indicate any new physics beyond the generic expectation. 
However, it has been argued~\cite{Gavin:2001uk, dccmodel} that a deviation from microscopic models like \textsc{hijing}~\cite{hijing} that include the decay of pions and incorporating realistic detector effects would be an ideal baseline to measure dynamical $\gc$ correlations beyond the generic expectation. 

Among other sources of dynamical correlation that might effect $\ndyn^{\gc}$, is the correlation coming from resonance decays. Similar type of dynamical correlation has been studied for the correlated production of kaons and pions by the STAR experiment~\cite{star_kpi, Tarnowsky:2012yk, Tribedy:2012qs}. It has been observed that when the correlated production of two species is dominated by resonance decays, $\ndyn$ will become negative. This particular behavior of $\ndyn$ was predicted in Ref.~\cite{nudyn}. A negative value of $\ndyn$ was also observed in correlation measurements of positively and negatively charged particles by STAR~\cite{star_chargecorr}. Resonance decays like $\rho^\pm \rightarrow \pi^\pm + \pi^0$ or $\omega \rightarrow \pi^0 + \pi^+ + \pi^-$, can introduce correlation between charged particles and photons. If $\ndyn^\gc$ is found to be non-zero, the sign of $\ndyn^{\gc}$ would indicate whether the dynamical correlation between photons and charged particles is dominated by resonances or not. For a comparative study, we measure the correlation of positively and negatively charged particles in the same acceptance of our analysis. 
Due to correlation coming from the decays of neutral resonances into
 a pair of charged particles, we expect $\ndyn^{\ch^+-\ch^-}$ to become negative.  

Several other measurement related effects influence the value of $\ndyn$. These include the event-by-event variation of acceptance due to variation of collision vertex position, event-by-event variations of efficiency, and the effect of mis-identification. We argue that the results for mixed events and $\textsc{geant}$ simulation using similar kinematic cuts used in case of data will help us to understand these effects. We also incorporate such effects in the estimation of systematic uncertainties as mentioned in a later section. 
}

{\color{black} Another variable, $r_{m,1}$, also called the robust variable, introduced by the MiniMax collaboration specifically for the search of DCC~\cite{Brooks:1999xy,Minimax} is defined as }
\bea
 \!\!\!\!\!\!r_{m,1}^\gc=\frac{\la N_\ch(N_\ch-1) \cdots \lf N_\ch-m+1\rf \, N_\ph \ra \nc}{\la N_\ch(N_\ch-1) \cdots (N_\ch-m)\ra \np}. 
\label{eq_rm1}
\eea
{\color{black} By construction, all the moments of $r_{m,1}$ are equal to unity for the Poissonian case, and higher order moments show a larger sensitivity to the (anti-)correlated signals.  This variable was also designed to remove explicit efficiency dependence~\cite{Brooks:1999xy, Minimax, dccmodel}}. It follows from Eq.(\ref{eq_ndyn}) and Eq.(\ref{eq_rm1}) that $r_{1,1}^\gc=\mathrm{corr}_{\gc}/\omega_\ch$. {\color{black} So, for the lowest order $m=1$, $\ndyn^\gc$ already includes all the information about $r_{m,1}$. The higher orders of $r_{m,1}$ will include additional sensitivity to any form of dynamical correlation as compared to $\ndyn^\gc$. There is one additional advantage of $r_{m,1}^\gc$ over $\ndyn$. As already mentioned above, the generic limit and the Poisson limits are not the same for $\ndyn$. It can be shown that for the generic case of pion production, $r_{m,1}$ becomes unity to all orders in $m$, which is also the Poisson limit of the observable. This makes it easier to interpret the results of this observable. This is due to the absence of the $\omega_\gamma$ term in $r_{m,1}$; the effects of additional fluctuations due to decay of neutral pions are absent in this observable~\cite{Brooks:1999xy, Minimax, dccmodel}.}
The functional dependence of $r_{m,1}$ on $m$ has been calculated 
in Refs.~\cite{dccmodel,Brooks:1999xy,Minimax,Mohanty:2005mv} { and is given by,}
\be
r_{m,1}^\gc=1-\frac{m \zeta}{(m+1)}, 
\label{eq_rm1sig}
\ee
{\color{black} where $0 \le |\zeta| \le 1$ is a parameter related to the strength of the $\gc$ (anti-)correlation. Positive values of $\zeta$ correspond to an anti-correlation and negative $\zeta$ corresponds to correlation. The value $\zeta=0$ corresponds to statistical fluctuations (Poisson limit). As aforementioned, the generic production also corresponds to $\zeta=0$, leading to $r_{m,1}^\textsc{gen}=1$ ~\cite{Brooks:1999xy,Minimax,dccmodel}. If charged and neutral particles are produced purely from the decay of DCC domains, $\zeta$ will become unity making $\rmn^\textsc{dcc}=1/(m+1)$~\cite{Brooks:1999xy,Minimax,dccmodel}.}
 
{\black It should be noted that like $\ndyn^{\gc}$, $\rmn^{\gc}$ is also not completely immune to sources of contaminations that introduce spurious correlations between charged hadrons and photons.
As previously stated, charged particles measured by FTPC have an impurity of 5$\%$, while the photons from PMD have an impurity of $30\%$. These impurities in the measurement of charged particles and photons affect the observables $\ndyn$ and $\rmn$. In reference to the collision vertex, the PMD is positioned after the FTPC. So the conversion photons that are detected as two charged particles in FTPC, are unlikely to reach PMD. Nevertheless some of the charged tracks, which are already identified by the FTPC, may fall on the PMD, which will give rise to spurious correlation. 
 We have tried to quantify such effects by comparing the values of these observables using a \textsc{geant} simulation~\cite{geant} with events from the \textsc{hijing} model (version 1.382), since event-by-event implementation of these effects by hand are not straightforward.}
 
  Throughout this analysis, we have studied the centrality dependence of the observables in terms of the experimental quantity $\ncp$, which represents the average multiplicity in the region of interest. We do so because this does not invoke any introduction of model dependence in our experimental results.  In many limiting scenarios, it has been shown that $\ndyn^\gc$ would become a function of $\ncp$ only. In the scenario of the aforementioned generic case of pion production, it can be shown~\cite{dccmodel} that  the observable $\ndyn^\gc$ will be proportional to $1/\ncp$.  An application of the Central Limit Theorem \cite{Trainor:2000dm, Luo:2010by} indicates that $\ndyn^\gc$ would show a $A+B/\sqrt{\langle N_\ch \rangle  \langle N_\ph \rangle}$ dependence on multiplicity \cite{dccmodel}, where $A$ and $B$ are constants related to the strength of the $\gc$ correlation. For a Boltzmann gas of pions in the grand canonical ensemble~\cite{ Begun:2004gs, Begun:2010ec}, one also predicts $\ndyn^\gc \sim1/\ncp$. For consistency, we use the same quantity to present the centrality dependence of $r_{m,1}$.

\subsection{Monte-Carlo Models}

{\color{black} We compare our results to the \textsc{hijing} model. In previous measurements by the STAR collaboration~\cite{Adams:2005aa, star-pmd-ftpc, Abelev:2009cy}, it was shown that \textsc{hijing} does a good job in describing the average multiplicity of charged particles and photons in the forward rapidity measured by the FTPC and PMD respectively. So it is expected to serve as a good baseline for charged and neutral particle fluctuations and also input for $\textsc{geant}$ simulations. \textsc{ampt} also provides good description of average multiplicities at forward rapidity~\cite{Adams:2005aa, star-pmd-ftpc, Abelev:2009cy}. It was also shown in Ref.~\cite{Abelev:2009cy} that the detector response to both \textsc{hijing} and \textsc{ampt} are almost similar. However due to violation of charge conservation present in the current version of \textsc{ampt}~\cite{Lin:2014uwa}, it is not clear whether this model can be used for the correlation analysis presented here. So, we restrict this analysis and compare our results to raw \textsc{hijing} and \textsc{hijing}+\textsc{geant} calculations only. It must be noted, that the observables $\ndyn^{\gc}$, $\rmn^{\gc}$ were estimated in Ref.~\cite{dccmodel} using three different Monte-Carlo models: \textsc{hijing}, \textsc{ampt}, and \textsc{u}r\textsc{qmd}~\cite{urqmd}. In all three cases, the observables are found to be similar within the statistical uncertainties over a wide range of multiplicity. A detailed discussion of the physics assumptions of the different models are beyond the scope of this paper; however we would like to point out that dynamical fluctuations arising from the domains of DCC formation are absent in these Monte-Carlo models.}
 
\subsection{Mixed event analysis}

A mixed event analysis provides a good baseline for this correlation analysis. By mixing tracks from different events, one can remove all sources of (anti-)correlations, although detector effects like overall efficiency, acceptance, etc. will still be present in the mixed event. 
However, any form of mis-identification (for instance FTPC tracks giving clusters in the PMD) that leads to spurious correlations will be absent in the mixed events.
\begin{figure}[h]
\centering
\includegraphics[height=6.5cm, width=6.5cm]{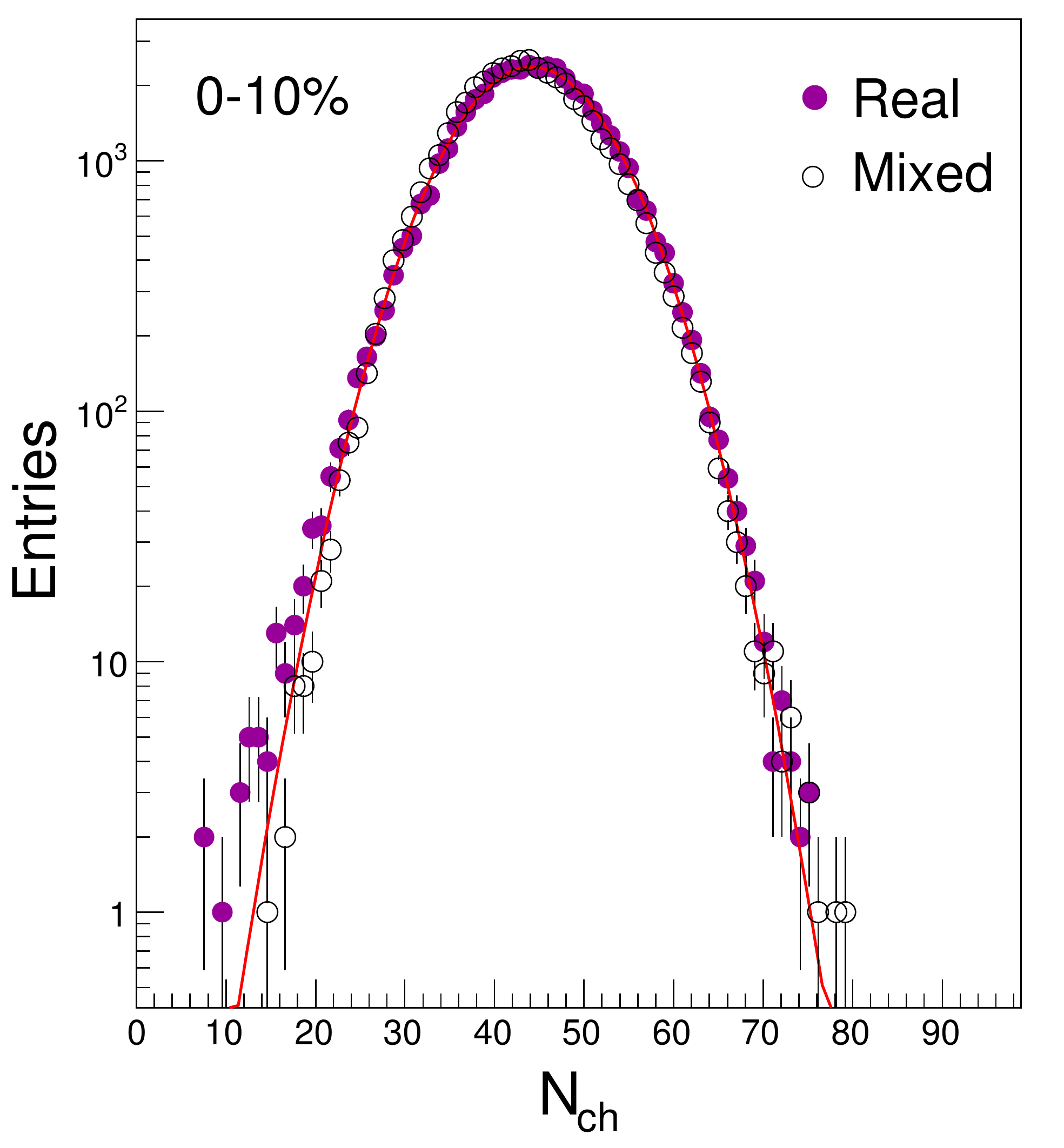} \\
\includegraphics[height=6.5cm, width=6.5cm]{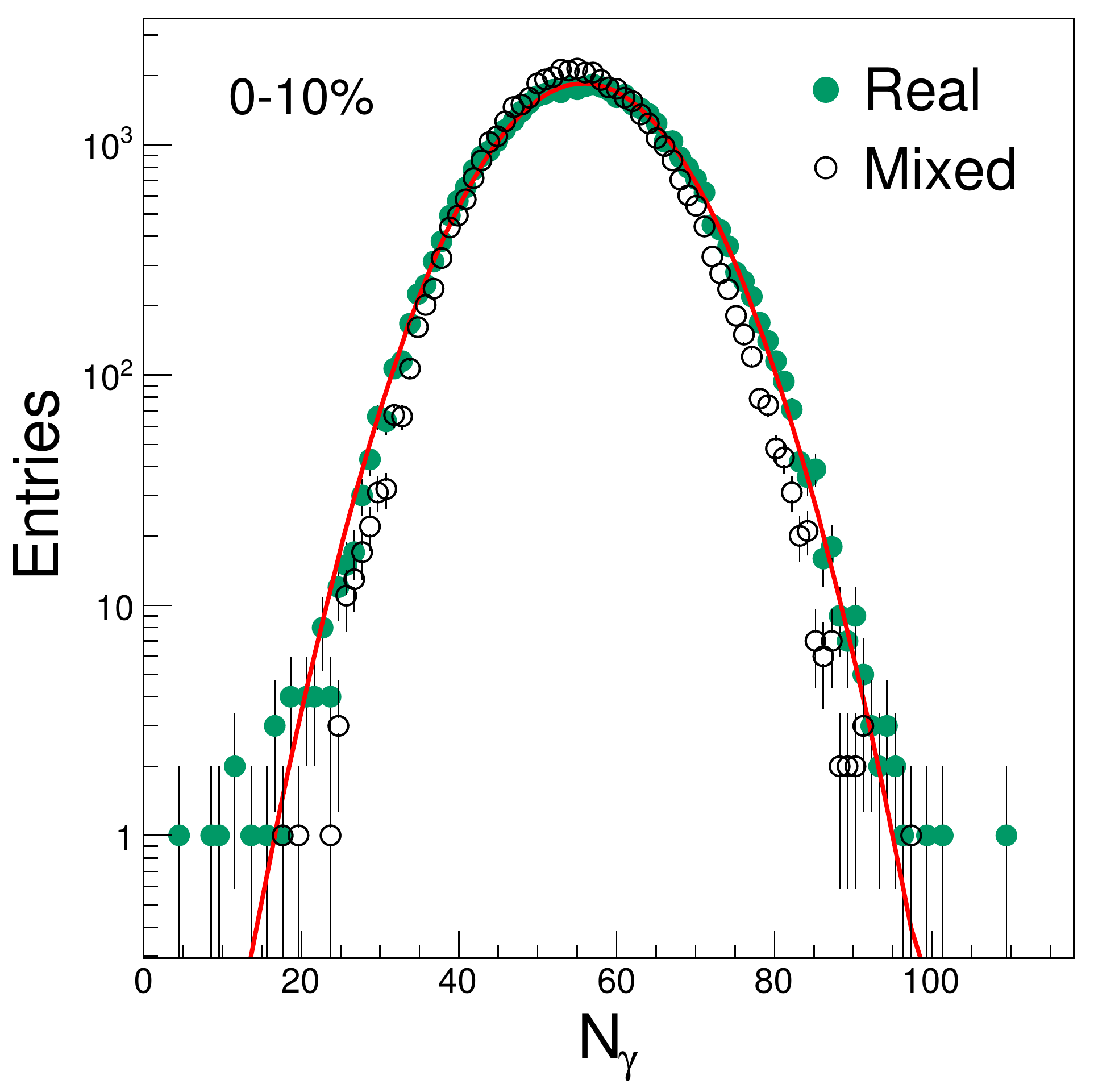}
\caption{\label{fig_mult} {\small (color online) Multiplicity distributions of raw charged particles and photons. The solid lines are Gaussian fits to the real event data points to guide the eyes.} }
\end{figure}
Mixed event sample construction consists of choosing a particular centrality bin and subdividing the events into narrow z-vertex bins. For a given real event, the total raw charged particle tracks from the FTPC and photon clusters from the PMD ($N_{\rm tot}=N_\ch^{\rm raw} + N_\ph^{\rm raw}$) are counted. 
In the next step, all the events that fall in the same z-vertex bin are scanned $N_{\rm tot}$ times to blindly pick up either a raw track or a cluster. In this way, a mixed event which has the same number of raw FTPC tracks + PMD clusters as the real event is constructed. Finally, all the kinematic cuts are applied and the total number of valid tracks, $N_{\ch}$, and valid photon clusters, $N_{\ph}$, are calculated.
 
To test the accuracy of our mixed event implementation, we show in Fig.\ref{fig_mult} the multiplicity distributions of charged particles and photons for the $0\!-\!10\%$ centrality bin for both real and mixed events. The real event distributions are fitted with a Gaussian curve to guide the eye. For peripheral centrality bins, the distributions can not be fitted by Gaussian distributions as they are not symmetric around the mean. We find that the real and mixed event distributions overlap reasonably well with the mean value of the distributions, agreeing within $2\%$. A similar trend is also observed for other centrality bins. The distributions shown in Fig.\ref{fig_mult} are for demonstration only and they are not corrected for detector efficiency. The purpose is to illustrate that the same effect of detector efficiency is also present in the mixed events. This would be useful in a later section to understand the response of the observables to detector efficiencies. It must be noted that due to random mixing of photons, the correlation among the two photons coming from the decay of a $\pi^0$ will also be missing in the mixed event samples. This reduces the fluctuation of photon numbers, resulting in a slightly narrower width of the $N_{\ph}$ distribution, which is visible in the lower panel of Fig.\ref{fig_mult}.

\subsection{Bin-width correction}
Bin-width effect is one of the most important corrections that need to be considered for any centrality-dependent event-by-event multiplicity fluctuation analysis~\cite{Luo:2013bmi, Adamczyk:2013dal, Adamczyk:2014fia}. This effect is a consequence of the fact that the centrality selection uses a distribution which is not flat. In this analysis, the centrality selection (event binning) is done using the distribution of the minimum bias multiplicity of charged particles in the mid-rapidity region, which is called the reference multiplicity distribution. The smallest possible centrality binning corresponds to dividing the distribution into every single value of the reference multiplicity. If a centrality bin is wide, it can correspond to a wide variation of the impact parameter (or system volume), that will propagate into the fluctuation of the final observable. Depending on the width of these bins, an artificial centrality dependence may be introduced in the final observable. This effect was demonstrated using the \textsc{u}r\textsc{qmd} model in Ref.\cite{Luo:2013bmi}. In order to correct this effect, event-by-event average quantities like the photon and the charged hadron multiplicities require weighted averages across the reference multiplicity distribution.
With the application of this correction, measures are independent of the chosen centrality bin width.
\begin{figure*}[htb]
\centering
\includegraphics[width=0.8\textwidth]{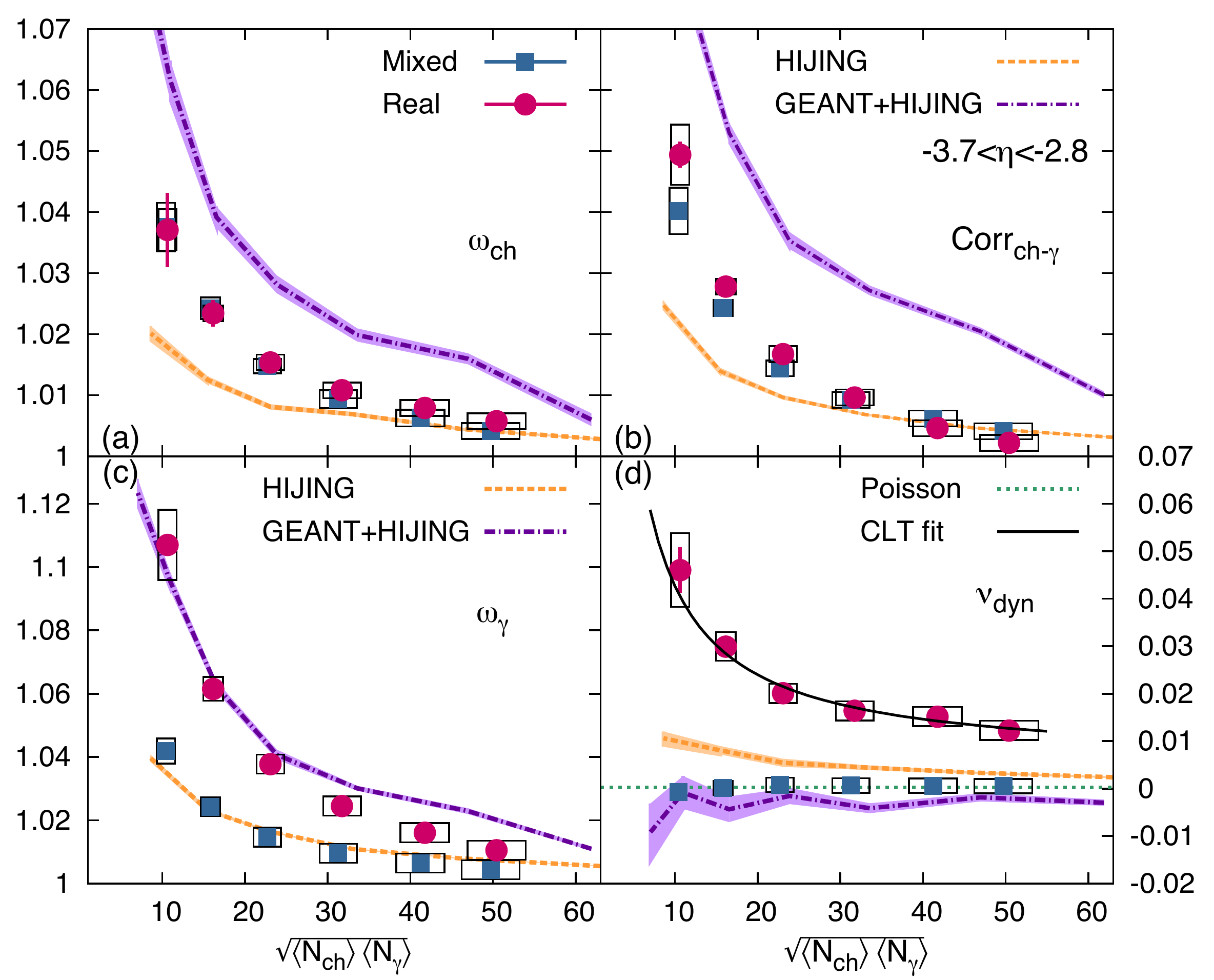}
\caption{\label{fig_ndynterms} (color online) The $\ndyn$ and the three terms of $\ndyn$ vs $\sqrt{ \langle N_\ch \rangle  \langle N_\ph \rangle}$ for real (circles) and mixed (squares) events. $\langle N_\ch \rangle$ and $\langle N_\ph \rangle$ are the mean charged particle and photon multiplicities in each centrality bin. Statistical uncertainties are represented by vertical lines while systematic uncertainties are shown by boxes. The statistical uncertainties for the models are shown by bands. The Central Limit Theorem (CLT) fit to $\ndyn$ for real events with the functional form $0.005+0.37/\sqrt{ \langle N_\ch \rangle  \langle N_\ph \rangle}$ is plotted as a solid curve.}
\end{figure*}
\subsection{Uncertainty analysis}
\subsubsection*{Statistical Uncertainty}
Calculations of statistical uncertainties have been performed using the bootstrap method~\cite{Efron:1979}. In the bootstrap method, (a) identical $n$ samples of minimum bias data sets are created by shuffling the event number. {\color{black} Each of these samples has the same number of events although the events are not identical, they will give rise to statistical variation of the observables.} (b) the bin width corrected observables $\ndyn$ and $r_{m,1}$ are calculated for each centrality bin separately for every event sample, and finally (c) estimated observables for these { $n$} different samples result in an approximately Gaussian distribution. The variance of this distribution is the statistical uncertainty from the bootstrap method. The number of samples $n$ is varied until the estimated uncertainty converges. For this analysis, we find that 100 samples provide good convergence. For the observable $\ndyn$, we have checked that the estimated uncertainty using the bootstrap method is consistent with the analytical error formula derived in ref.~\cite{errnudyn}.
\subsubsection*{Systematic Uncertainty}
{\color{black} Systematic uncertainties of $\ndyn$ and $\rmn$ are obtained by varying different quality cuts shown in Table~\ref{tab_datacuts}} on charged tracks and photon clusters. 
The variation of the maximum distance of closest approach of a track to the primary vertex by 0.5 cm introduces $\sim8\%$ variation in the value of the observables. The effect of possible charged particle contamination in the photon sample has been included in the systematic uncertainties. 
The systematic uncertainty from the charged hadron contamination is obtained by varying the cut (MIP cut) for photon-hadron discrimination discussed in Section \ref{data_reco}. Variation by one unit of MIP cut causes a $\sim 6 \%$ variation of the value of $\ndyn$. The variation of the primary collision vertex position in the z-direction induces an 8$\%$ variation of the observable $\ndyn$. The overall systematic uncertainty of $\ndyn$ is estimated to be $\sim 15 \%$ within the centrality range of $0-60\%$. Similar cuts were applied to evaluate the systematic uncertainty of the quantity $\sqrt{ \langle N_\ch \rangle  \langle N_\ph \rangle}$ and is estimated to be $\sim 7 \%$. The systematic uncertainty for the observable $r_{m,1}$ is estimated to be in the range of $0.001-0.002$ for value of $m=1\!-\!3$. The quantity relevant for the observable $r_{m,1}$ is its deviation from the generic limit. This systematic uncertainty of $r_{m,1}$ is found to be approximately $10\!-\!20\%$ of the magnitude of its variation of the observable $r_{m,1}$ from its generic limit. Systematic uncertainty for different observables are listed in Table~\ref{tab_syserr} in the Appendix.
We have investigated any possible affect of azimuthal correlations, such as anisotropic flow, on $\gc$ correlation. Connection between elliptic flow ($v_2$) and anomalous neutral pion production in the context of DCC like domain formation was first discussed in Ref.~\cite{Asakawa:2002vj}. Using events 
from $\textsc{hijing}$, we introduce $v_2$ by changing the azimuthal angle of each pion. We find that both the observable $\ndyn$ and $\rmn$ are insensitive to $v_2$ over a realistic range of values ($v_2=0\!-\!5\%$) at the forward rapidity~\cite{Back:2004zg}.
\section{Results}
\label{results}
{ Figures~\ref{fig_ndynterms}(a)-(c)} show the multiplicity (centrality) dependence of the different terms of $\ndyn$ in Eq.\ref{eq_ndyn} for real and mixed events. All three terms approach their respective Poisson limits (= 1) for higher values of multiplicity. The individual scaled fluctuation terms $\omega_\ch$ and $\omega_\ph$ (shown in Fig.~\ref{fig_ndynterms}(a) and Fig.~\ref{fig_ndynterms}(c))  are higher for real events compared to mixed events showing presence of additional non-statistical fluctuation in the data. This is also seen in Fig.~\ref{fig_mult} (lower panel) for the distribution of photons. 
The discrepancy between real and mixed events is much larger for the photon fluctuation term than for the charged particle fluctuation term. This is because in addition to the common origin of the multiplicity fluctuations from the parent particles (charged and neutral pions), the decays from neutral pions to photons introduce further fluctuations. This particular feature of the data is also consistent with the $\textsc{hijing}$ model calculation.
The scaled correlation term $\corr$ in Eq.\ref{eq_ndyn} is shown in Fig.~\ref{fig_ndynterms}(b). $\corr$ for real events when compared to the mixed events, the baseline, is larger in peripheral bins, comparable for the mid-central events and smaller in more central events. 
{\color{black}However, the statistical significance of the difference at high centrality is too small to draw any firm conclusion.}
We see similar trends with multiplicity ($\sqrt {\langle N_\ch \rangle  \langle N_\ph \rangle}$) for all the three terms. {\color{black} A close look at Fig.~\ref{fig_ndynterms}(a-c) indicates that the mixed-events give rise to a universal curve for all three terms. As discussed before, mixed events are supposed to include statistical fluctuations only, due to finite multiplicity they show large deviation from the Poisson limit (=1) of each term. Only towards most central events it becomes closer to unity. These individual curves also include similar measurement related artifacts (efficiency and acceptance) as the real event curves.}
\begin{figure}[htb]
\includegraphics[width=0.5\textwidth]{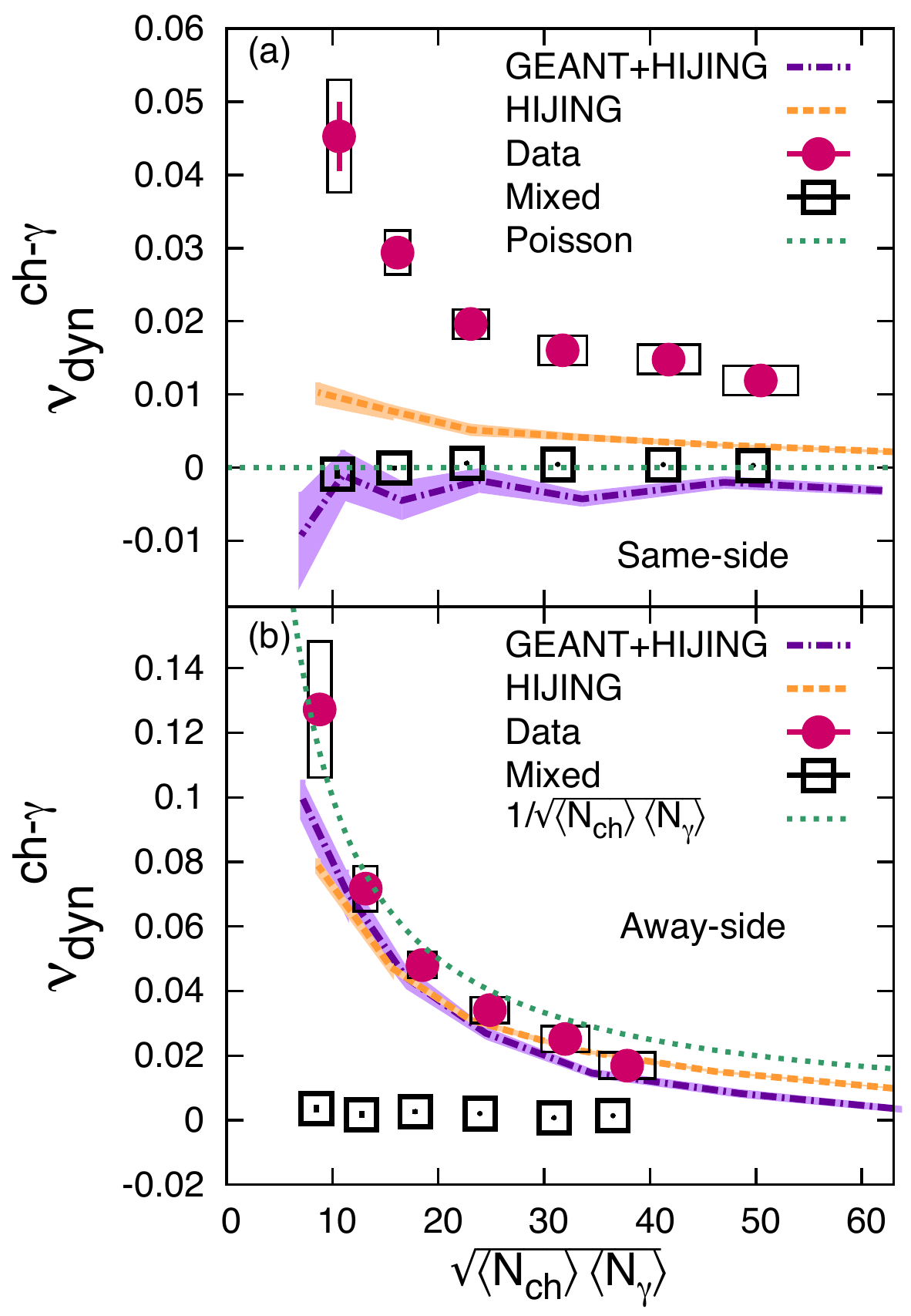}
\caption{\label{fig_ndynEW} (color online) The values of the observable $\ndyn^{\ch-\ph}$ for the same side (upper panel) and the away-side (lower panel). The same-side corresponds to a measurement of photons and charged particles in the same acceptance -3.7$<\eta<$-2.8. The away-side corresponds to measurement of photons in the range -3.7$<\eta<$-2.8 and charged particle in the range 3.7$>\eta>$2.8. Model and mixed event calculations are performed using same kinematic cuts. For data, the statistical uncertainties are shown by vertical lines and the systematic uncertainties are shown by boxes. For model curves statistical uncertainties are shown by bands.}
\hspace{5pt}
\end{figure}
\begin{figure}[htb]
\includegraphics[width=0.5\textwidth]{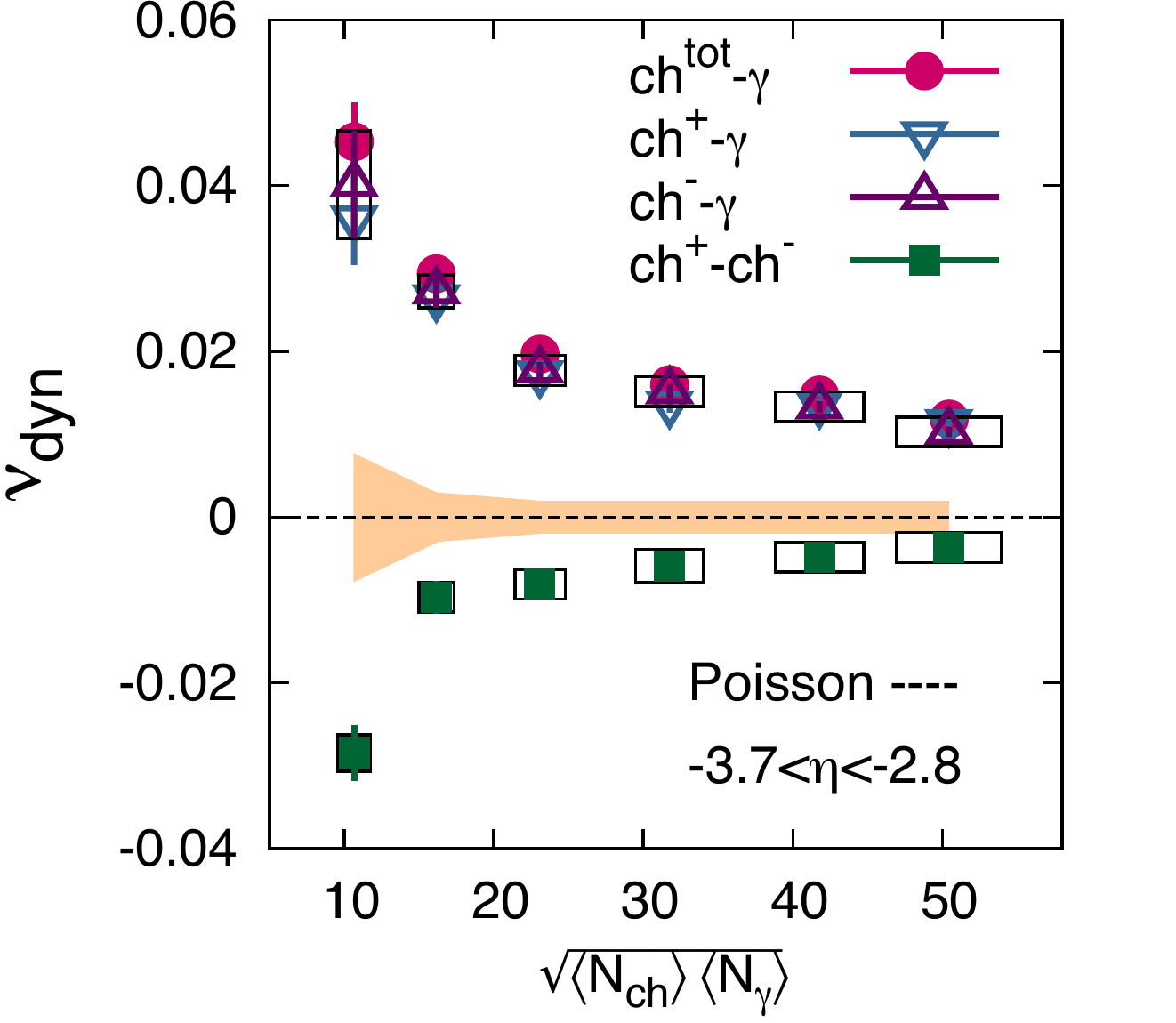}
\caption{\label{fig_ndyn_chdep} (color online) The correlation between positive and negative charged particles measured by the FTPC and photons measured by the PMD using $\ndyn$ for real events. For  the positive-negative charged particle correlation, the statistical uncertainties are represented by lines and the systematic uncertainties are shown by boxes. For the different combinations of $\gc$ correlation the systematic uncertainties in $\ncp$ are shown by boxes whereas the systematic uncertainties in $\ndyn$ are shown by the yellow band.}
\end{figure}
Fig.\ref{fig_ndynterms}(d) shows the variation of $\ndyn^\gc$ with $\sqrt {\langle N_\ch \rangle  \langle N_\ph \rangle }$ for real and mixed events. {\color{black} For the mixed events, the result is consistent with the Poisson expectation at all centralities. This demonstrates the interesting property of the observable $\ndyn$, which by construction eliminates the statistical fluctuations, and detector effects like efficiency and acceptance.}
For the real events, we see a non-zero positive value of $\ndyn^\gc$. We fit the data points for real events with a function of the form $A+B/\sqrt {\langle N_\ch \rangle  \langle N_\ph \rangle}$ as per CLT predictions~\cite{dccmodel}. {\color{black} The values of the parameters $A$ and $B$ are found to be 0.005 and 0.37 respectively. The fit quality is not very good since we find a $\chi^2/\mathrm{dof}\sim 2$. However, within the error bars, we do not see any significant deviation from the CLT prediction.}

In the same plot, we show \textsc{hijing} and \textsc{hijing}+\textsc{geant} results for comparison. The value of $\ndyn$ is very close to the Poisson limit for \textsc{hijing} in more central events, with a positive value that shows a similar trend as the data. Results from \textsc{hijing} events simulated through \textsc{geant} are also close to the Poisson expectation within statistical uncertainties. {\color{black} We see small difference between the \textsc{hijing} and the \textsc{hijing}+\textsc{geant} curves. We argue that this is due to the spurious correlation coming from mis-identification of photons that can not be eliminated even by the construction of the observable $\ndyn$. The difference between the \textsc{hijing} curve and the \textsc{hijing}+\textsc{geant} curve serves as a reference to how much this detector effect is still present in the data sample that can not be excluded from the presented analysis. It must be noted that this detector effect does not change the conclusion that the observed value of $\ndyn^\gc$ is positive, since the contamination has the opposite effect to the deviation seen in data. For the present measurement, it is evident that the model curve shows very small deviations from the Poisson curve compared to data. Data show non-zero positive values for all centrality bins, indicating the presence of dynamical fluctuation for all centralities. 

The measurement of $\ndyn^\gc$ shown in Fig.\ref{fig_ndynterms} comes from charged particles and photons in the same pseudo-rapidity range of $-3.7\!<\!\eta\!<\!-2.8$ (same-side). In Fig.\ref{fig_ndynEW} we compare this result with the $\ndyn^\gc$ using photons measured in the pseudo-rapidity range of $-3.7\!<\!\eta\!<\!-2.8$ and charged particles measured in the pseudo-rapidity range of $2.8\!<\!\eta\!<\!3.7$ (away-side). The area in the $\eta-\phi$ space for photons and charged particles is the same in both cases, however due to lower reconstruction efficiency of charged tracks in the range $2.8\!<\!\eta\!<\!3.7$ for the second FTPC, we show the data points only up to $\ncp\approx 38$. Fig.\ref{fig_ndynEW}(a) shows that, for the same side, a large difference between data and model curves is observed. For the away-side (Fig.\ref{fig_ndynEW}(b)) the difference between data and model curve almost disappears. 
This strengthens our argument that the difference between data and model observed in the same side is not due to detector effects. A closer look at Fig.\ref{fig_ndynEW} indicates that the small difference between the \textsc{hijing} and \textsc{hijing}+\textsc{geant} curve in the same side disappears in case of away-side. This indicates that the effect of contamination (that brings down the absolute value $\ndyn$) is absent in the away-side and $\ndyn$ is robust enough to eliminate all other measurement related artifacts. In the away-side, data and model seem to follow a universal trend,  which is dependent only on the value of $\ncp$, more specifically very close to a value of $1/\ncp$ as shown by a dotted curve on the same plot. Keeping Fig.\ref{fig_ndynEW} (b) as a reference, one can argue that the deviation seen in the same side is of dynamical origin. It should be noted that the absolute value of $\ndyn^\gc$ has gone up in the away-side, which is evident in both data and models. For \textsc{hijing}, this growth corresponds to an increase of $\ndyn$ from almost zero (same-side) to something close to $1/\ncp$ (away-side). This is due to the decrease of the correlation term (${\rm corr}_\gc$) over a rapidity unit of about 3.2. A possible explanation of this can be found in Ref.~\cite{nudyn}. It has been argued that the multiplicity correlation function drops like a Gaussian with increasing relative pseudo-rapidity $\Delta\eta$. The width of the Gaussian is dependent on centrality~\cite{nudyn}, the effect of which is also present in $\textsc{hijing}$. Since the correlation term of Eq.\ref{eq_ndyn} drops with $\Delta\eta$, the absolute value of $\ndyn$ goes up with $\Delta\eta$. The mixed event points are consistent with the Poisson limit as expected. Since by construction any form of correlation is eliminated, the drop of correlation with $\Delta\eta$ is not relevant for mixed events results.}  
\begin{figure}[tb]
\vspace{20pt}
\includegraphics[width=0.5\textwidth]{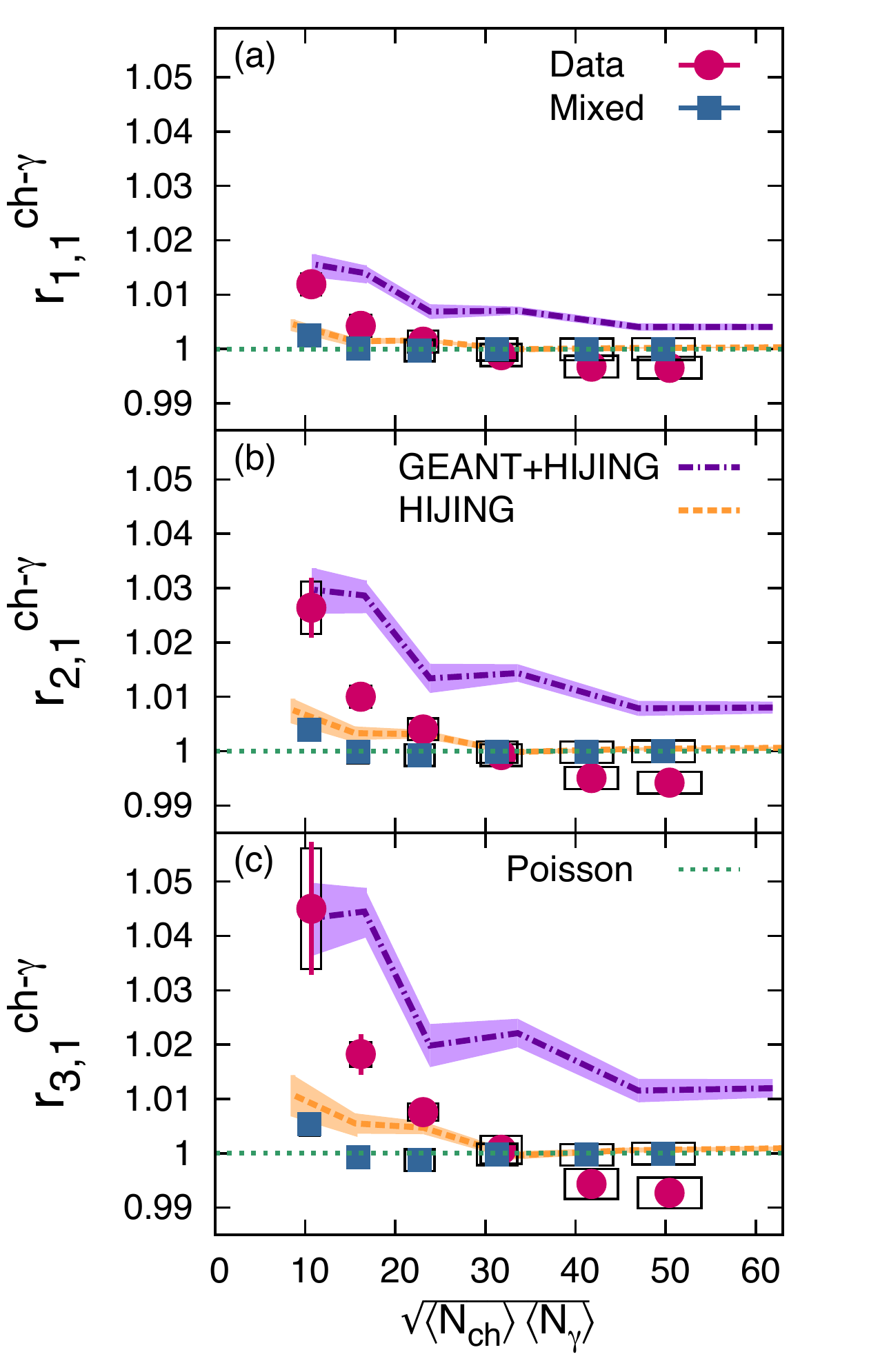}
\caption{\label{fig_rm1} (color online) $r_{m,1}$ vs multiplicity for first three orders of $m$. Data and mixed event results are compared to {\textsc{hijing} and \textsc{hijing}+\textsc{geant}} values, which are shown by the curves. Statistical uncertainties are shown by vertical lines and the systematic uncertainties are shown by boxes. For model curves statistical uncertainties are shown by bands.}
\end{figure}
\begin{figure}[htb]
\vspace{20pt}
\includegraphics[width=0.5\textwidth]{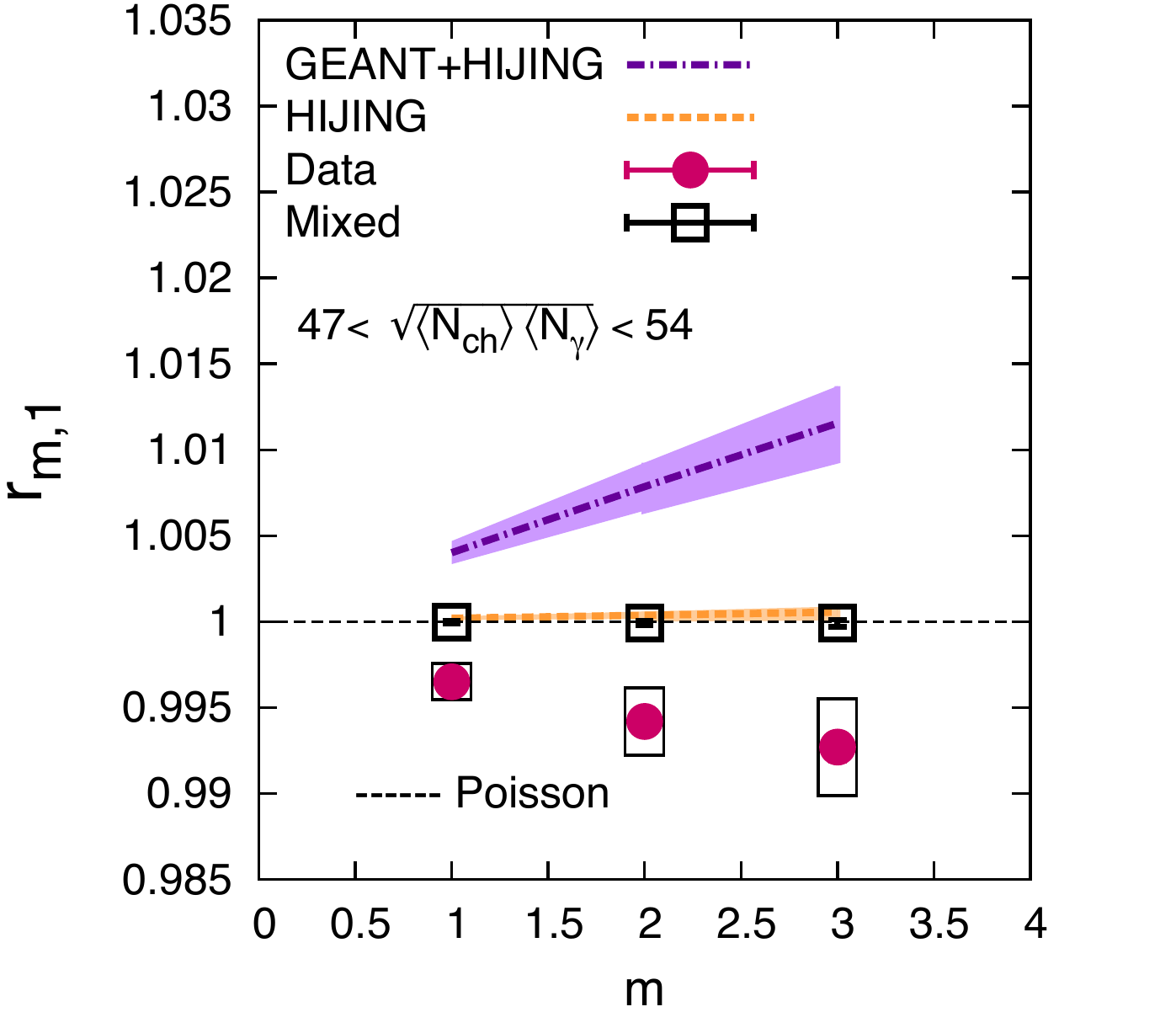}
\caption{\label{fig_rm1m} (color online)  The moments $r_{m,1} (m\!=\!1\!-\!3)$ for real and mixed events as a function of order $m$ in a fixed multiplicity bin of $47\!<\!\ncp\!<\!54$. {\textsc{hijing},  \textsc{hijing}+\textsc{geant}} results are represented by curves. For data the statistical uncertainties are shown by vertical lines and the systematic uncertainties are shown by boxes. For model curves statistical uncertainties are shown by bands.}
\end{figure}
{\color{black} In order to investigate the possible origin of the dynamical fluctuation seen in $\ndyn^\gc$, we study the charge dependence of $\ndyn$. As shown in Fig.\ref{fig_ndyn_chdep}, results for combinations of photons with individual (positive and negative) charges are very close to that of photon and total charged particle correlation. 
The results for the combination of positively and negatively charged particles is very different in sign and magnitude compared to that of $\gc$ correlation.
The observable $\ndyn$ for positively and negatively charged particles is negative. The reason is that it is dominated by the large correlation term that arises from pairs of oppositely charged particles produced from the decay of neutral resonances. 
This result is consistent with the previous measurement by STAR at mid-rapidity in Au+Au collision at $\sqrt{s_{NN}}=200$ GeV~\cite{star_chargecorr}. A simple model of resonance production studied in Ref. \cite{nudyn} indicates that for particle production dominated by the decay of resonances, the observable $\ndyn$ will be negative.
The fact that the correlation pattern between photons and charged particles is opposite to that of negative and positive particles would indicate a different production mechanism.
So the current measurement, as shown in Fig.~\ref{fig_ndyn_chdep}, unambiguously supports the conclusion that the dynamical fluctuation observed in case of $\ndyn^\gc$ is not dominated by  resonance decay effects. The results thus indicate that a completely different mechanism is responsible for the correlated production of charged particles and photons.}

The nature and strength of the $\gc$ correlation is further explored using the observable $\rmn$. This observable was designed to study deviations from a generic pion production scenario that would correspond to a value of unity. {\color{black} Fig.\ref{fig_rm1} shows the variation of the first three moments $r_{m,1}~(m\!=\!1\!-\!3)$ with $\ncp$ for real and mixed events.} The \textsc{hijing} and \textsc{hijing}+\textsc{geant} curves are also displayed. $r_{m,1}$ is nearly constant with $\ncp$ for both \textsc{hijing} and mixed-events. Multiplicity dependence (with $\ncp$) of $r_{m,1}$ shows that the data points are lower than the mixed-events, \textsc{hijing}, and \textsc{hijing}+\textsc{geant} in central collisions.
 We see that the mixed event results are consistent with the generic (Poisson) limit of the observables. Raw $\textsc{hijing}$ values are also very close to the generic limit. This could indicate that the correlated production of pions in $\textsc{hijing}$ is very similar to that of generic production. %
 However, the data deviate from this trend. A similar trend is observed in case of \textsc{hijing}+\textsc{geant}, but the values of $r_{m,1}$ are always greater than unity. {\color{black} However, we only see $<\sim1\%$ deviation from the Poisson limit. Passing $\textsc{hijing}$ events through $\textsc{geant}$ changes $r_{m,1}$ in a direction opposite to that seen in the data. This difference between $\textsc{hijing}$ and \textsc{hijing}+\textsc{geant} for $r_{1,1}$ is about $1\%$ which is consistent with a similar difference observed for $\ndyn$ in Fig.\ref{fig_ndynterms}. 
 For higher orders of $r_{m,1}$, the difference between $\textsc{hijing}$ and \textsc{hijing}+\textsc{geant} increases up to $1-3\%$ over the entire range of multiplicity. The origin of this is the contamination present in the data sample. For real events (data), $r_{m,1}$ dips slightly below unity for higher multiplicities, showing a small deviation from the generic case. For values of $\ncp$ greater than 30, this would correspond to $\zeta < 0.01$ in Eq.(\ref{eq_rm1sig}).}
   
  Since we expect higher orders of $r_{m,1}$ to be more sensitive to any deviation from the generic limit, we plot in Fig.\ref{fig_rm1m} the variation of $r_{m,1}$ with $m$ in the window of multiplicity $47<\ncp<54$. The moments $r_{m,1}$ show with $m$ in real data a trend which is opposite to that in \textsc{hijing}, \textsc{hijing}+\textsc{geant} simulation, and mixed events. However, for each order, we see the deviation from the generic case to be small and lies within a range of $0.99-1$. 

\section{summary $\&$ conclusion}
\label{summary}
Correlations between photon and charged particle multiplicities at $-3.7<\eta<-2.8$ have been measured in STAR using the PMD and the FTPC in Au+Au collisions at $\sqrt{s_{NN}}$=200 GeV. The observables $\ndyn$ and $r_{m,1}$ have been used as measures of correlation. Measured $\ndyn$ {from data shows a non-zero, positive value that exceeds the predictions from \textsc{hijing}, mixed events, and \textsc{hijing}+\textsc{geant} when charged particles and photons are measured in the same acceptance. When charged particles are measured in a different acceptance ($3.7>\eta>2.8$) compared to photons ($-3.7<\eta<-2.8$), the difference between model prediction and data is found to be negligible. This indicates the presence of dynamical fluctuations in the ratio of inclusive charged to photon multiplicities. $\ndyn$ shows an approximate $1/\ncp$ dependence as expected from the Central Limit Theorem. The charge dependence of $\ndyn^{\gamma-\rm ch}$ shows that different combinations of $\gc$ correlations are alike, but behave differently (both in magnitude and sign) when compared to $\ndyn^{\ch^+-\ch^-}$ obtained for the combination of positive and negatively charged particles in the same acceptance. This indicates that the mechanism of correlated production of oppositely charged particles is different from the correlated production of neutral and charged particles and, at the same time, the $\gc$ correlation is not dominated by correlations from decays. A second observable $\rmn$, also called the Minimax observable, is used to extract any deviation of $\gc$ correlation from the expectation of generic pion production. The centrality dependence of $r_{m,1}~(m\!=\!1\!-\!3)$ shows a different trend compared to that from mixed events and \textsc{hijing}. $r_{m,1}$ is below the generic (or Poisson) limit at higher multiplicity. For central events, $\rmn$ {as a function of the order $m$ shows a trend opposite to that from models, suggesting very small deviation from the expectation of the generic production of pions. {\color{black} For all the orders the deviation is found to be less than $1\%$ from the generic expectation.} Additional exploration of the origin of this deviation by quantitative estimation is beyond the sensitivity of our measurement.} 
\clearpage
\begin{widetext}
\section*{appendix}
\begin{table}[h]
\caption{\label{tab_syserr} List of total systematic uncertainties for the observables used in this analysis.}
\centering
\begin{tabular}{lllll}
\hline
&{\small $\surd{\!\langle N_\ch \rangle \! \langle N_\ph \rangle} $}& Err$(\surd{\!\langle N_\ch \rangle \! \langle N_\ph \rangle})$ & $\ndyn^\text{real}$ & Err($\ndyn^\text{real}$)\\
\hline
\hline
& 50 &4 &0.012 &0.002 \\
& 42 &3 &0.015 &0.002 \\
& 32 &2 &0.016 &0.002 \\
& 23 &2 &0.019 &0.001 \\
& 16 &1 &0.029 &0.003 \\
& 10.7 &0.8 &0.045 &0.008 \\
\hline
\hline
\end{tabular}
\hspace{5pt}
\begin{tabular}{lllll}
\hline
&{\small $\surd{\!\langle N_\ch \rangle \! \langle N_\ph \rangle} $}& Err$(\surd{\!\langle N_\ch \rangle \! \langle N_\ph \rangle})$ & $\ndyn^\text{mixed}$ & Err($\ndyn^\text{mixed}$)\\
\hline
\hline
&50 &3 &0.0003 &0.0002 \\
&41 &3 &0.0004 &0.0003 \\
&31 &2 &0.0004 &0.0003 \\
&23 &2 &0.0006 &0.0003 \\
&16 &1 &-0.0001 &0.0006\\ 
&10.5 &0.8 &-0.001 &0.001\\
\hline
\hline
\end{tabular}
\label{tab_syserr}
\vspace{10pt}
\centering
\begin{tabular}{lllll}
\hline
&{\small $\surd{\!\langle N_\ch \rangle \! \langle N_\ph \rangle} $}& Err$(\surd{\!\langle N_\ch \rangle \! \langle N_\ph \rangle})$ & $r_{1,1}^\text{real}$ & Err($r_{1,1}^\text{real}$)\\
\hline
\hline
&50 &4  &0.997 &0.001 \\ 
&42 &3  &0.997 &0.001 \\ 
&32 &2  &0.999 &0.001 \\ 
&23 &2  &1.001 &0.001 \\ 
&16 &1  &1.004 &0.001 \\ 
&10.6 &0.8&1.012 &0.002 \\
\hline
\hline
\end{tabular}
\hspace{5pt}
\begin{tabular}{lllll}
\hline
&{\small $\surd{\!\langle N_\ch \rangle \! \langle N_\ph \rangle} $}& Err$(\surd{\!\langle N_\ch \rangle \! \langle N_\ph \rangle})$ & $r_{1,1}^\text{mixed}$ & Err($r_{1,1}^\text{mixed}$)\\
\hline
\hline
&50 &3 &0.9999 &0.0001 \\ 
&41 &3 &0.9999 &0.0002 \\ 
&31 &2 &0.9999 &0.0003 \\ 
&23 &2 &0.9997 &0.0004 \\ 
&16 &1 &1.0000 &0.0003 \\ 
&10.6 &0.8 &1.0025 &0.0005 \\
\hline
\hline
\end{tabular}
\label{tab_syserr}
\vspace{10pt}
\centering
\begin{tabular}{lllll}
\hline
&{\small $\surd{\!\langle N_\ch \rangle \! \langle N_\ph \rangle} $}& Err$(\surd{\!\langle N_\ch \rangle \! \langle N_\ph \rangle})$ & $\ndyn^{\ch\pm}$ & Err($\ndyn^{\ch\pm}$)\\
\hline
\hline
&50 &4 &-0.0037 &0.0006\\ 
&42 &3 &-0.005 &0.001 \\ 
&32 &2 &-0.006 &0.002 \\ 
&23 &2 &-0.008 &0.001 \\ 
&16 &1 &-0.0096 &0.0009 \\ 
&10.7 &0.8 &-0.028 &0.002 \\ 
\hline
\hline
\end{tabular}
\hspace{5pt}
\begin{tabular}{lllll}
\hline
&m& Err$(m)$ & $r_{m,1}^\text{real}$ & Err($r_{m,1}^\text{real}$)\\
\hline
\hline
&1 &0 &0.997 &0.001 \\ 
&2 &0 &0.994 &0.002 \\ 
&3 &0 &0.993 &0.003 \\ 
\hline
\hline
\end{tabular}\\
\hspace{5pt}
\begin{tabular}{lllll}
\hline
&{\small $\langle N_\ph \rangle $}& {\small Err$\langle N_\ph \rangle$} & {\small $\langle N_\ch \rangle $}  & {\small Err$\langle N_\ch \rangle$}\\
\hline
\hline
& 57 & 7 &44 & 3 \\ 
& 45 & 6 &39 & 2 \\ 
& 32 & 4 &32 & 2 \\ 
& 22 & 3 &24 & 1 \\ 
& 15 & 2 &17.6 & 0.6 \\ 
& 10 & 1 &11.9 &0.4 \\ 
\hline
\hline
\end{tabular}
\label{tab_syserr}
\end{table}
\end{widetext}
%
%
\section*{acknowledgement}
We thank the RHIC Operations Group and RCF at BNL, the NERSC Center at LBNL, the KISTI Center in Korea, and the Open Science Grid consortium for providing resources and support. This work was supported in part by the Offices of NP and HEP within the U.S. DOE Office of Science, the U.S. NSF, CNRS/IN2P3, FAPESP CNPq of Brazil,  the Ministry of Education and Science of the Russian Federation, NNSFC, CAS, MoST and MoE of China, the Korean Research Foundation, GA and MSMT of the Czech Republic, FIAS of Germany, DAE, DST, and CSIR of India, the National Science Centre of Poland, National Research Foundation (NRF-2012004024), the Ministry of Science, Education and Sports of the Republic of Croatia, and RosAtom of Russia.


\begin{thebibliography}{50}
%
%
\bibitem{Adams:2005dq} 
  J.~Adams {\it et al.}  [STAR Collaboration],
  Nucl.\ Phys.\ A { 757}, 102 (2005) 
  [nucl-ex/0501009].

\bibitem{Back:2004je} 
  B.~B.~Back {\it et al.} [PHOBOS Collaboration],
  Nucl.\ Phys.\ A { 757}, 28 (2005)
  [nucl-ex/0410022].
  
\bibitem{Adcox:2004mh} 
  K.~Adcox {\it et al.}  [PHENIX Collaboration],
  Nucl.\ Phys.\ A { 757}, 184 (2005)
  [nucl-ex/0410003].
  

\bibitem{Arsene:2004fa} 
  I.~Arsene {\it et al.}  [BRAHMS Collaboration],
  Nucl.\ Phys.\ A { 757}, 1 (2005)
  [nucl-ex/0410020].



\bibitem{Jeon:1999gr}
  S.~Jeon and V.~Koch,
  Phys.\ Rev.\ Lett.\  { 83}, 5435 (1999)

  
\bibitem{Bjd} J.D. Bjorken, What lies ahead?, SLAC-PUB-5673, 1991.


\bibitem{Blaizot:1992at}
  J.~P.~Blaizot and A.~Krzywicki,
  Phys.\ Rev.\  D { 46}, 246 (1992).

\bibitem{Rajagopal:1992qz}
  K.~Rajagopal and F.~Wilczek,
  Nucl.\ Phys.\  B { 399}, 395 (1993)

\bibitem{Rajagopal:1995bc}
 K.~Rajagopal,
 arXiv:hep-ph/9504310.

\bibitem{Adams:2005aa} 
  J.~Adams {\it et al.}  [STAR Collaboration],
  Phys.\ Rev.\ Lett.\  { 95}, 062301 (2005)
  
\bibitem{Randrup:1996es}
  J.~Randrup,
  Nucl.\ Phys.\  A { 616}, 531 (1997)

\bibitem{Randrup:1997kt}
  J.~Randrup and R.~L.~Thews,
  Phys.\ Rev.\  D { 56}, 4392 (1997)

\bibitem{Randrup:1996ay}
  J.~Randrup,
  Phys.\ Rev.\ Lett.\  { 77}, 1226 (1996).
 
\bibitem{Rajagopal:2000yt}
  K.~Rajagopal,
  Nucl.\ Phys.\  A { 680}, 211 (2001)


\bibitem{Asakawa:all} 
  M.~Asakawa, H.~Minakata and B.~Muller,
  Nucl.\ Phys.\ A { 638}, 443C (1998),Phys.\ Rev.\ C { 65}, 057901 (2002).
  


\bibitem{Gavin:1993bs} 
  S.~Gavin, A.~Gocksch and R.~D.~Pisarski,
  Phys.\ Rev.\ Lett.\  { 72}, 2143 (1994)
  [hep-ph/9310228].

\bibitem{Gavin:1993px} 
  S.~Gavin and B.~Muller,
  Phys.\ Lett.\ B { 329}, 486 (1994)
  [hep-ph/9312349].

\bibitem{Gavin:1995cp} 
  S.~Gavin,
  Nucl.\ Phys.\ A { 590}, 163C (1995).
  
\bibitem{Gavin:2001uk} 
  S.~Gavin and J.~I.~Kapusta,
  Phys.\ Rev.\ C { 65}, 054910 (2002)
  [nucl-th/0112083].

\bibitem{Bellwied:1998yf} 
  R.~Bellwied, S.~Gavin and T.~Humanic,
  nucl-th/9811085.


\bibitem{Aggarwal:1997hd}
  M.~M.~Aggarwal {\it et al.}  [WA98 Collaboration],
  Phys.\ Lett.\  B { 420}, 169 (1998)

\bibitem{Aggarwal:2000aw}
  M.~M.~Aggarwal {\it et al.}  [WA98 Collaboration],
  Phys.\ Rev.\  C { 64}, 011901 (2001)
  
\bibitem{Aggarwal:2002tf}
  M.~M.~Aggarwal {\it et al.}  [WA98 Collaboration],
  Phys.\ Rev.\  C { 67}, 044901 (2003)

\bibitem{Collaboration:2011rsa}
  M.~M.~Aggarwal {\it et al.},
  Phys.\ Lett.\  B { 701}, 300 (2011)

\bibitem{Brooks:1999xy}
  T.~C.~Brooks {\it et al.}  [MiniMax Collaboration],
  Phys.\ Rev.\  D { 61}, 032003 (2000)
 
 \bibitem{Minimax}
  T.~C.~Brooks {\it et al.}  [MiniMax Collaboration],
  Phys.\ Rev.\  D { 55}, 5667 (1997)
  
  
\bibitem{star_nim}
               K.~H.~Ackermann {\it et al.}, 
               Nucl. Instr. Meth. A 499, 624 (2003).

\bibitem{startpc_nim}
	     M. Anderson {\it et al.}, 
	     Nucl. Instr. Meth. A 499, 659 (2003)
	       
	       
\bibitem{starbemc_nim}
M.~Beddo {\it et al.}, 
  Nucl. Instr. Meth. A 499, 725 (2003)

  
\bibitem{starpmd_nim} 
               M.~M.~Aggarwal {\it et al.}, 
               Nucl. Instr. Meth. A 499, 751 (2003);
               M.~M.~Aggarwal {\it et al.}, 
               Nucl. Instr. Meth. A 488, 131 (2002).

  
\bibitem{starftpc_nim} 
               K.~H.~Ackermann  {\it et al.}, 
               Nucl. Instr. Meth. A 499, 713 (2003).
               

               
\bibitem{nudyn}
  C.~Pruneau, S.~Gavin and S.~Voloshin,
  Phys.\ Rev.\  C { 66}, 044904 (2002)
  [arXiv:nucl-ex/0204011].


 \bibitem{dccmodel}
 P.~Tribedy {\it et al.} 
   Phys.\ Rev.\  C { 85}, 024902 (2012).

  
\bibitem{star-pmd-ftpc} 
  J.~Adams {\it et al.}  [STAR Collaboration],
  Phys.\ Rev.\ C { 73}, 034906 (2006)
  [nucl-ex/0511026].
  
  
\bibitem{Luo:2013bmi} 
  X.~Luo, J.~Xu, B.~Mohanty and N.~Xu,
  J.\ Phys.\ G { 40}, 105104 (2013)
  [arXiv:1302.2332 [nucl-ex]].

\bibitem{Voloshin:1999yf} 
  S.~A.~Voloshin, V.~Koch and H.~G.~Ritter,
  Phys.\ Rev.\ C { 60}, 024901 (1999)
  [nucl-th/9903060].

\bibitem{Wang:2012bga} 
  Q.~Wang and F.~Wang,
  arXiv:1205.4638 [nucl-ex].


  
\bibitem{star_kpi}
  B.~I.~Abelev {\it et al.}  [STAR Collaboration],
  Phys.\ Rev.\ Lett.\  { 103}, 092301 (2009)
  [arXiv:0901.1795 [nucl-ex]].

\bibitem{star_dcc}
  S.~M.~Dogra  [STAR Collaboration],
  J.\ Phys.\ G { 35}, 104094 (2008).

\bibitem{Tribedy:2012qs} 
  P.~Tribedy [STAR Collaboration],
  Nucl.\ Phys.\ A904-905, 463c (2013)
  [arXiv:1211.0171 [nucl-ex]].
  
    
\bibitem{errnudyn} 
  P.~Christiansen {\it et al.}
 Phys.\  Rev.\  C { 80}, 034903 (2009).	
  [ arXiv:0902.4788 [hep-ex]].
  
\bibitem{hijing}
  X.~N.~Wang and M.~Gyulassy,
  Phys.\ Rev.\ D { 44}, 3501 (1991).
  
  
\bibitem{Tarnowsky:2012yk} 
  T.~J.~Tarnowsky [STAR Collaboration],
  Acta Phys.\ Polon.\ Supp.\  { 5}, 515 (2012)
  [arXiv:1201.3336 [nucl-ex]].
  
  \bibitem{star_chargecorr} 
  B.~I.~Abelev {\it et al.}  [STAR Collaboration],
  Phys.\ Rev.\ C { 79}, 024906 (2009)
  [arXiv:0807.3269 [nucl-ex]].
  
\bibitem{Mohanty:2005mv}
 B.~Mohanty and J.~Serreau,
 Phys.\ Rept.\  { 414}, 263 (2005)
  [arXiv:hep-ph/0504154].
  
    \bibitem{geant}
  V.~Fine and P.~Nevski, in Proceedings of CHEP-2000, Padova, Italy, p.\ 143. 
   
  
\bibitem{Trainor:2000dm}
  T.~A.~Trainor,
  [hep-ph/0001148].
  
\bibitem{Luo:2010by}
  X.~F.~Luo, B.~Mohanty, H.~G.~Ritter and N.~Xu,
  J.\ Phys.\ G { 37}, 094061 (2010)
  [arXiv:1001.2847 [nucl-ex]].
  

\bibitem{Begun:2004gs}
  V.~V.~Begun, M.~Gazdzicki, M.~I.~Gorenstein and O.~S.~Zozulya,
  Phys.\ Rev.\  C { 70}, 034901 (2004)
  [arXiv:nucl-th/0404056].
\bibitem{Begun:2010ec}
  V.~V.~Begun, M.~I.~Gorenstein and O.~A.~Mogilevsky,
  Phys.\ Rev.\  C { 82}, 024904 (2010)
  [arXiv:1004.2918 [nucl-th]].





\bibitem{Abelev:2009cy} 
  B.~I.~Abelev {\it et al.}  [STAR Collaboration],
  Nucl.\ Phys.\ A { 832}, 134 (2010)
  [arXiv:0906.2260 [nucl-ex]].
  
  
\bibitem{Lin:2014uwa} 
  Z.~W.~Lin, Acta Phys. Polon. Supp. 7, 191 (2014)
  arXiv:1403.1854 [nucl-th],
  Z.~W.~Lin, C.~M.~Ko, B.~A.~Li, B.~Zhang and S.~Pal,
  Phys.\ Rev.\  C { 72}, 064901 (2005)
  [arXiv:nucl-th/0411110].



\bibitem{urqmd}
  S.~A.~Bass {\it et al.},
  Prog.\ Part.\ Nucl.\ Phys.\  { 41}, 255 (1998)
  [Prog.\ Part.\ Nucl.\ Phys.\  { 41}, 225 (1998)]
  [arXiv:nucl-th/9803035].
  


  
\bibitem{Adamczyk:2013dal} 
  L.~Adamczyk {\it et al.}  [STAR Collaboration],
  Phys.\ Rev.\ Lett.\  { 112}, 032302 (2014)
  [arXiv:1309.5681 [nucl-ex]].
  
  
  
\bibitem{Adamczyk:2014fia} 
  L.~Adamczyk {\it et al.}  [STAR Collaboration],
  arXiv:1402.1558 [nucl-ex].
  

  
\bibitem{Efron:1979}
  Bradley Efron,
 SIAM Review\ 
 Vol.\ 21, No.\ 4 (1979), pp. 460-480.

  
 \bibitem{Asakawa:2002vj} 
  M.~Asakawa, H.~Minakata and B.~Muller,
  Nucl.\ Phys.\ A { 721}, 305 (2003)
  [nucl-th/0212070].
  
\bibitem{Back:2004zg} 
  B.~B.~Back {\it et al.}  [PHOBOS Collaboration],
  Phys.\ Rev.\ Lett.\  { 94}, 122303 (2005)
  [nucl-ex/0406021].
  


  
 
\end{thebibliography}
\end{document}